# Competing mechanisms govern the thermal rectification behavior in semi-stochastic polycrystalline graphene with graded grain-density distribution

Simanta Lahkar*, Raghavan Ranganathan

*Department of Materials Engineering, Indian Institute of Technology Gandhinagar, Gujarat 382355, India*

**Abstract**

Thermal rectifiers are devices that have different thermal conductivities in opposing directions of heat flow. The realization of practical thermal rectifiers relies significantly on a sound understanding of the underlying mechanisms of asymmetric heat transport, and two-dimensional materials offer a promising opportunity in this regard owing to their simplistic structures together with a vast possibility of tunable imperfections. However, the in-plane thermal rectification mechanisms in 2D materials like graphene having directional gradients of grain sizes have remained elusive. In fact, understanding the heat transport mechanisms in polycrystalline graphene, which are more practical to synthesize than large-scale single-crystal graphene, could potentially allow a unique opportunity, in principle, to combine with other defects and designs for effective optimization of thermal rectification. In this work, we investigate the thermal rectification behavior in periodic atomistic models of polycrystalline graphene whose grain arrangements were generated semi-stochastically to have different gradient grain-density distributions along the in-plane heat flow direction. We employ the centroidal Voronoi tessellation technique to generate realistic grain boundary structures for graphene, and the non-equilibrium molecular dynamics simulations method is used to calculate the thermal conductivities and rectification values. Additionally, detailed phonon characteristics and propagating phonon spatial energy densities are analyzed based on the fluctuation-dissipation theory to elucidate the competitive interplay between two underlying mechanisms, namely, (i) propagating phonon coupling and (ii) temperature-dependence of thermal conductivity that determines the degree of asymmetric heat flow in graded polycrystalline graphene.

**Introduction**

Developing technologies that can improve energy sustainability is a major priority for our society in the current times. New and efficient thermal devices can play a central role in this regard in a variety of applications. A thermal rectifier is one such thermal nanodevice that preferentially conducts heat in one direction, analogous to its electronic counterpart, and can potentially open up novel technological frontiers of applications such as continuous thermoelectric renewable-power generation, thermal logic circuits, and thermal management systems for atomic clocks if adequate efficiencies can be achieved[1,2]. Compared to bulk materials, where asymmetric heat flow characteristics can be achieved through a limited number of ways, two-dimensional (2D) materials possess a wide range of tunable structural features that can create a suitable structural asymmetry affecting the heat transfer efficiencies in the two opposite directions with respect to the direction of asymmetry. Graphene is a particularly promising 2D material for this technology owing to its extremely high in-plane thermal conductivity[3]. Some of these strategies to tune the thermal rectification (TR) behavior include features like defects, doping, functionalization, grain density, interfacial structures, geometries, and boundary conditions[2,3,4,5,6,7]. However, their practical realization relies on a robust understanding of the underlying heat transport phenomena[1,2].

In our earlier work, we found a cross-plane thermal rectification mechanism in out-of-plane gradient hydrogen-functionalized multilayer graphene (G-MLG), which was attributed to the phonon density of states (PDOS) mismatch between the heat source and sink [4]. In general, the mechanisms that are responsible for causing asymmetric heat flow behavior in 2D materials can broadly be divided into two kinds: 1) difference between the phonons produced at the hotter end of the device for heat flow in forward and reverse directions, respectively, that can propagate across the length of the device with varying efficiencies affecting the relative ease of heat flow between the two directions, and 2) disparate temperature-sensitivities of the local thermal conductivities in the different sections of the device close to either end that result in different average conductivity depending on the location of the heat source and sink[3]. A common strategy of TR of the first kind involves asymmetric interfacial structures between two materials having different phonon transport properties. This was the attributed mechanism for the up to 44% thermal rectification (% difference in thermal conductivities between the two directions) predicted by Liu et al. at the silicene/graphene interface[5]. In our previous study, we observed tunable cross-plane TR in multilayer graphene with gradient hydrogenation in consecutive layers through molecular dynamics (MD) simulations attributed to a primary mechanism of the first kind[4]. Geometrical features like triangular holes in a graphene sheet are also expected to cause asymmetric phonon scattering for different directions of heat flow[8]. Another study by Wang et al. used MD simulations to show that having a tapered width or a doping of carbon nanoparticle in one end of graphene sheets led to a prominent scattering of the propagating phonons when the narrow or doped end of the sheet is at the heat source, causing a TR of over ~10% which they had observed experimentally in micrometres-sized graphene sheets[3]. However, the rectification due to such mechanisms should reduce with increasing sample length as the heat transport shifts farther from the ballistic to diffusive regime due to a reduction in the strength of coupling between the phonons across the device length. Wang et al. also observed a 26% TR experimentally in graphene sheets with an asymmetric nanopore distribution owing to the second kind of thermal rectification. Here, the thermal conductivity in the end without nanopores had a high negative dependence on the local temperature, whereas the side with pores had almost temperature-independent thermal conductivity, resulting in a difference in the net thermal conductivity of the rectifier upon exchange of the hot and cold sides[3]. This approach of rectification was further studied in detail by Hu et al., who illustrated theoretically that introducing regular nanopores pattern in pristine graphene to create a graphene phononic crystal (GPnC) resulted in higher phonon localization and, consequently, caused a reduction in the temperature ($T$) dependence of thermal conductivity, $k$, ($k \propto T^{-\alpha}$ and $\alpha \approx 1$ in pristine graphene)[9]. Interestingly, the degree of reduction in α depends on the nanopore and patterned-domain size, which prompted Hu et al. to add more than one GPnCs with different pore sizes in a series circuit with pristine graphene on one side leading to an increase in the TR[10].

While the majority of research focus in this field has so far been on engineering single crystalline 2D rectifiers, it is important to elucidate the effects of polycrystalline structures on the rectification mechanisms in order to realize practical 2D thermal rectifiers at length scales beyond several micrometres, owing to the difficulty associated with producing and processing pristine, large area, single crystalline 2D materials. Though a recent study using non-equilibrium molecular dynamics simulations (NEMD)[11] found TR in polycrystalline graphene with a special design of grain width gradient across the device[12], our ability to properly understand and design such a polycrystalline graphene thermal rectifier is still at a very nascent stage, critically due to a lack of understanding and evidence of the underlying mechanisms of TR in general grain-size or grain-density graded polycrystalline graphene, or any other 2D material, in the literature. Here, we report a systematic study of TR in *stochastically generated* atomistic models of polycrystalline graphene sheets having a gradient distribution of grain densities across the length of the layers and uncover the interplay between two simultaneously acting distinct mechanisms that cause the asymmetric heat flow in opposite directions in such graded polycrystalline graphene (GPG) structures.

**Simulation and analysis methods**

In order to obtain realistic atomistic models of polycrystalline 2D materials, the central Voronoi tessellation (CVT) algorithm[13] was used to relax the atomistic structure of grain boundaries and the lattice around it to be in-line with experimental observations. Classical non-equilibrium molecular dynamics (NEMD) simulations (using the LAMMPS[14] package) were used to study the steady-state heat flow behaviour in the different models as it provides the required balance between accuracy and scale to study the relevant thermal phenomena effectively[11]. The steady-state local temperature of each atom during the simulation was calculated as:

$$T_i = \frac{\langle m_i v_i \cdot v_i \rangle}{3k_B} \qquad (1)$$

where $m_i$ and $v_i$ are the mass and velocity of the atom $i$ and $k_B$ is the Boltzmann constant. The simulation starts by repeated minimization with in-plane relaxation and MD equilibration runs in a canonical ensemble (NVT) until the pressure and potential energy of the system upon (Gaussian) randomized velocity initiation fall below a set threshold. This is followed by gradual heating and cooling of the respective heat baths for 500 ps, and then maintaining the set bath temperatures for the rest of simulation, where the atoms outside the heat baths are in a micro-canonical (NVE) ensemble. Once steady state has been reached the calculation of the average temperature profiles and the heat flow rate is done by averaging over at least 1.3 ns or more, depending on the nature of the simulation, which is used to calculate the thermal conductivities based on the temperature difference between the heat source and the heat sink in the structure, $\Delta T$[11], using the following relationship:

$$k = \frac{Q.L}{A.\Delta T} \qquad (2)$$

Where $Q/A$ is the steady state heat flux, and $L$ is the distance between the heat source and the heat sink. The value of $\Delta T \approx 1.35 \times L$ is used to maintain good signal to noise ratio and the linear transport regime, where temperature and length units are K and nm, respectively. The atoms in two narrow 0.5 nm edges at the two ends along the direction of heat flow are kept fixed during the heat transfer simulation. A timestep of 0.2 fs is used for the simulations to have sufficient accuracy. The Langevin thermostat, shown not to cause any unphysical phonon density of states characteristics, was used to model the heat source and sink in the simulations[11]. The width of the temperature baths were made sufficiently large in comparison to the damping to ensure proper thermalization of the phonons in the simulation[11]. If the thermal conductivities of a structure in the two opposite directions are $k_1$ and $k_2$, respectively, in the order of decreasing magnitude, the TR coefficient can be calculated as:

$$TR = \frac{(k_1 - k_2) \times 100}{k_2} \qquad (3)$$

The optimized Tersoff forcefield developed by Kinaci et al. has been used to describe the graphene inter-atomic interactions[15] owing to its high accuracy in thermal analysis of graphene using MD simulations[16]. The fluctuation-dissipation theory, which can capture the anharmonic effects of temperature on phonons, was used to analyze the phonon localization their spatial energy distribution, using the phonon package implemented in LAMMPS that calculates the dynamical matrix for a structure from MD simulations[17]. The participation ratio of different phonon modes[3] can be calculated as:

$$P_\lambda = 1 \Big/ N \sum_i \left( \sum_\alpha \varepsilon^*_{i\alpha,\lambda} \varepsilon_{i\alpha,\lambda} \right)^2 \qquad (4)$$

where $\varepsilon_{i\alpha,\lambda}$ is the $\alpha$th eigenvector component of the phonon mode $\lambda$ with frequency $\omega_\lambda$ for the $i$th atom and $N$ is the total number of atoms. The $\varepsilon_{i\alpha,\lambda}$ and $\omega_\lambda$ can be calculated by diagonalizing the dynamical matrix[18]. These values were used, together with the atomic structure and per atom temperature information, to map the the spatial energy distribution of propagating phonons on atom $i$ given by:

$$E_i = \sum_\omega \sum_\lambda \sum_\alpha \left( n + \frac{1}{2} \right) \hbar \omega \varepsilon^*_{i\alpha,\lambda} \varepsilon_{i\alpha,\lambda} \delta(\omega - \omega_\lambda) \qquad (5)$$

where $n$ is the phonon occupation number given by the Bose-Einstein distribution, and $\lambda$ spans all the long range phonons with $P_\lambda$ greater than a set threshold[3].

**Results and discussion**

We first designed a protocol to create the atomistic structures of GPG having realistic and randomized distributions of grain shapes and orientations across the devise, while also possessing an average arrangement of grain densities across the GPG structure that resembled any arbitrary user-defined design, or layout. We started with an arbitrary layout of grain densities at different points along the length of the structure, which will guide the creation of grains that follow any desired distribution, including a uniform or a monotonically increasing/decreasing grain number density, along the length of the structure. This layout is then mirrored to represent a periodic distribution in the 2D space. Then a set of regular points is generated based on the layout of number densities from the previous step that are distributed in a rectangular region representing the size of the final periodic atomistic GPG structure. Following this step, random perturbations are introduced to the set of regular points in the form of vectorial displacements for each point that follow a Gaussian distribution with the mean at 0 the standard deviation being scaled to 1/8$^{th}$ of the inter-point distance based on the corresponding number density that was used to create the given point. Thus, the degree of perturbations in this step can be changed, which affects the statistical deviations of the final atomistic polycrystalline graphene structure versus a regular arrangement of grains that very closely follows the prescribed grain density distribution layout without sufficient realistic randomness in grain shapes and orientations. Then the points are used as the generators to create a Voronoi tessellation whose edges represent the grain boundary regions and each polygon in the tessellation is used to generate a graphene lattice having 0.246 nm lattice constant[19] and random orientations. Finally, the centroidal Voronoi tessellation algorithm[13] is used to create a realistic grain boundary structure having only pentagon and heptagon pairs per experimental observations of polycrystalline graphene. This scheme of semi-stochastic polycrystalline structure creation for graphene is illustrated in Fig. 1.

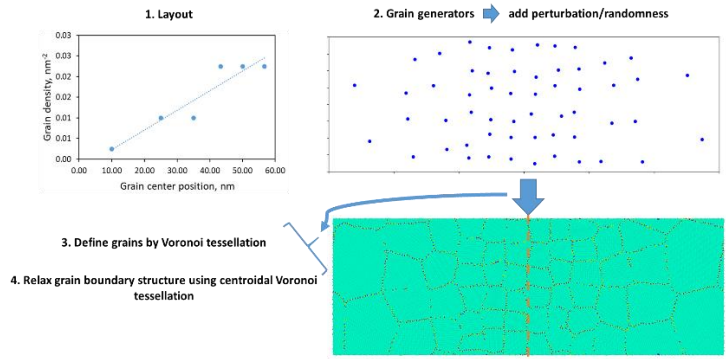

**FIG. 1.** Semi-stochastic atomistic modeling scheme for graded polycrystalline graphene. The initial layout is mirrored and added to create a periodic layout which is used to define the regular grain generator positions. The dotted line at the middle of the periodic atomistic polycrystalline graphene model shows the location at which the model is split to generate two structures each having a monotonic change in average grain density along its length.

Subsequently, we define a new parameter to characterize the key aspect of the degree of the variation of grain density across the length of the structures, denoted as the grain density gradient parameter, which is calculated as the difference between the maximum and minimum values of grain densities divided by the distance between the respective regions in the unit GPG structure as described above. In this study, all the GPG structure have a linear variation of the number of grains per unit width across the length of the structure. Once the periodic GPG models are created, it is divided exactly in half and each half, which has the maximum and minimum grain densities at its two ends, is used as a separate model for the atomistic conductivity simulations as shown in Fig. 2.

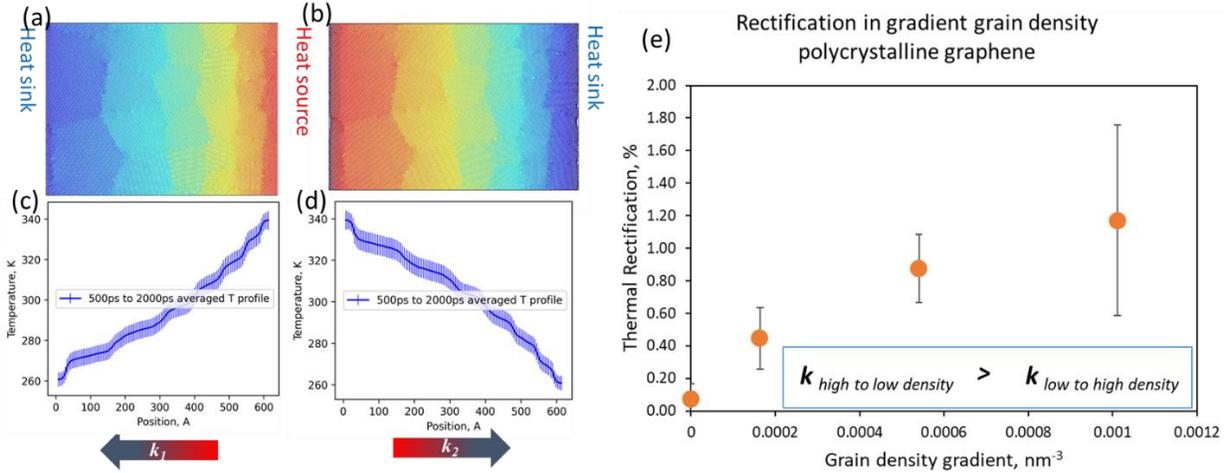

**FIG. 2.** NEMD calculation and thermal rectification in GPG. (a) and (b) are atomic snapshots for the conductivity simulations in opposite directions with respect to the grain density gradient, with atoms colored based on the per atom temperatures averaged through the interval of the simulation used for conductivity calculation (the range of the color bar is from 258 K for blue and 342 K for red). (c) and (d) plot the temperature profile for (a) and (b), respectively, averaged over the same duration with the standard deviation (SD) as error bars. (e) shows the average percent TR for different GPG structures with the SD as error bars calculated from the different values obtained by repeating the conductivity simulations with different starting velocities. The first point along x-axis corresponds to single crystal graphene.

We use non-equilibrium molecular dynamics (NEMD) simulations[11] to determine the thermal conductivities, $k$'s, and the corresponding TRs for the different atomistic graphene models. In this technique, where the temperature difference is first imposed using different thermostats (heat baths) at the source and sink regions (Fig. 2(a) and (b)) and the resultant heat flux is used to calculate the thermal conductivity after a steady state has been reached, which is characterized by the development of a steady temperature profile in the structure (Fig. 2(c) and (d)) and the convergence between the heat input and output by the source and sink region thermostats respectively [11].

Fig. 2(e) plots the TR values obtained for single crystalline graphene and GPG models having different grain density gradients the similar simulation box length of about 100 nm along the heat transport direction. The TR plot clearly indicates that the $k$ is larger when heat is flowing from high grain density to low grain density region in GPG, with practically zero TR for single crystalline graphene as expected. The direction of TR observed from NEMD simulations in this study agree with the observations in a previous study on the NEMD TR simulations of polycrystalline graphene with linear variation of grain width along the length[12]. Besides the direction of the TR observed, the relationship between the grain density gradient and the TR value in GPG also indicates that the magnitude of TR gets higher with increasing grain density gradient in the structure, i.e., the TR increases with increasing difference between the terminal grain densities (in other words, the terminal grain sizes) for the same length of the GPG structure.

Next, we try to understand the underlying mechanism of TR in GPG by focusing on the phonon characteristics in polycrystalline graphene at any arbitrary temperature. Fig. 3 compares the phonon localization characteristics in single crystalline graphene versus a representative uniform grain-size polycrystalline graphene model, calculated using the fluctuation-dissipation theory[20], which is implemented as the phonon package in LAMMPS and calculates the dynamical matrix for a structure from MD simulations[17, 21, 22], in order to understand the underlying mechanism of asymmetric heat transport property. Since we are interested in analyzing polycrystalline samples, we use the full-sized periodic atomistic structure for both single crystal and polycrystal graphene as a single unit cell without any replication for our simulations and phonon analysis, and then compare the relative differences between the different structures having different average grain sizes. In order to obtain reliable phonon properties we equilibrated the structure as NVT ensemble, after repeated minimization and relaxation as described above, and ran the simulation for over 1.3 ns to calculate the dynamical matrix[18]. This method of calculating the phonon properties can also capture the anharmonic effects of temperature on phonons[17].

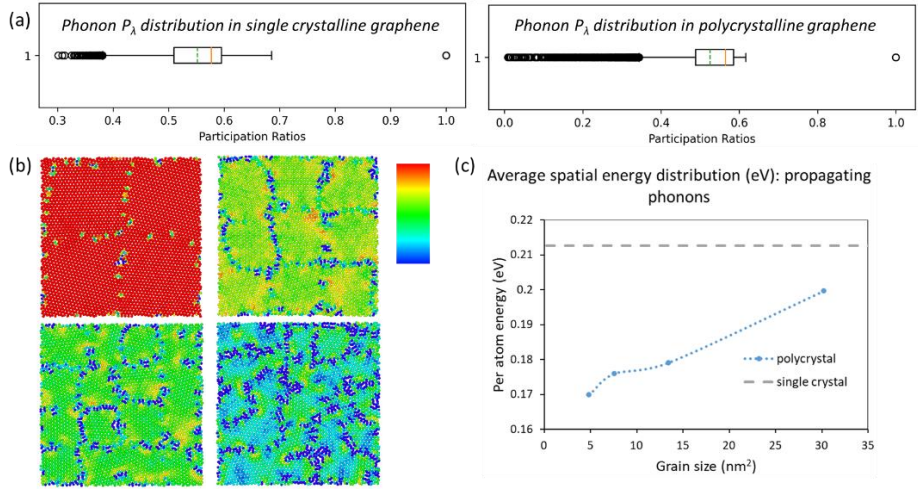

**FIG. 3** Phonon Localization and spatial energy distribution of propagating phonons in single crystalline and polycrystalline graphene. (a) shows two box plots illustrating the respective statistical distributions of the $P_\lambda$'s of phonons in a single crystal and a polycrystalline graphene (with average grain size of 13.44 nm$^2$). The orange and green dotted lines are at the positions of the medians and modes of the respective distribution. The ends of the box denotes the 25$^{th}$ and 75$^{th}$ percentile points in the distribution, while ends of the whiskers are at the last data point within a distance of 1.5*IQR from the respective ends of the box (IQR = 75$^{th}$ percentile – 25$^{th}$ percentile values) and the circles denote the position of the outliers. The increased number of outliers with low $P_\lambda$, the higher proximity of the mean to the 25$^{th}$ percentile point, and the significant shortening of the upper whisker clearly indicates the increased phonon localization and reduction in the range of propagating phonons in polycrystalline graphene as compared to single crystalline graphene. (b) shows the map of spatial energy distribution of propagating phonons in 4 different uniform grain sized graphene models (the upper red and lower blue ends of the color bar correspond to the energy of 0.19 and 0.165 eV/atom respectively), whose average values are plotted in (c) versus corresponding average grain size joined by a dotted line along with another dotted line at the value of the average spatial energy distribution for single crystal graphene.

The participation ratio, $P_\lambda$, of a phonon gives the proportion of atoms in the unit sample that contribute to the given phonon mode, which is a measure of the range of the phonons. Thus the long range phonons will have a value of $P_\lambda$ much closer to 1 while the short range (localized) phonons will have a $P_\lambda$ closer to 0. From Fig. 3 (a), it can be clearly seen that polycrystalline graphene has significantly greater localization of phonons, as well as a reduction in the range of long range phonons, as compared to the single crystalline graphene. The long range phonons, also known as the propagating phonons, are primarily responsible for thermal conduction in graphene[23]. We can map the energy carried by the propagating phonons in a given structure by summing up the spatial energy distributions for all the propagating phonons with participation ratio over a certain threshold.

Thus, we created different uniform grain size polycrystalline graphene models with the same total unit cell size of 11 nm X 11 nm and calculated the spatial energy distribution of propagating phonons (having $P_\lambda > 0.4$) for these structures (Fig. 3(b)) equilibrated at the same temperature of 300 K. The corresponding variation of the mean values of propagating phonon spatial energy distributions for different grain sizes of polycrystalline graphene is plotted in Fig. 3(c), along with the corresponding values for continuous single crystal graphene sheets. We find that the propagating phonons carry much more energy in large grain sizes, i.e., for low grain densities, with single crystal graphene having the maximum energy of propagating phonons. Further, an increase in temperature also leads to an increase in the average spatial energy distribution values for all the grain sizes due to a significant increase in the phonon occupation number, *n*, in equation (5), which we have also verified. This clearly indicates that if the large grain size region is at the heat source, then propagating phonons with higher energy will be produced at the heat source, as compared to the converse case when the smaller grain size region is at the heat source. This should cause the heat transport to be more efficient from the low grain density to high grain density region in GPG[3]. We further verified that for a fully periodic GPG structure (where the two heat baths were placed respectively at the start and middle of the structure) with a length of about 20 nm and width of about 7 nm (Fig. 4), the average spatial energy distribution of propagating phonons is higher for the graded structure when the heat flows from region with low grain density (larger grain size) to the region with high grain density (smaller grain size). This could be attributed to the higher energy propagating phonons originating from the low grain density region at the heat source, which couple across the structure increasing the energy of propagating phonons at the other end as well, and consequently increases the heat flow efficiency of the whole structure in that direction. This is indeed observed from the increased thermal conductivity measured in this particular direction for this size of GPG structure. However, this does not align with the observed direction of TR in GPG of larger length, i.e., 100 nm and above (Fig. 2(e)). Hence, the higher energy propagating phonon coupling should *not* be the primary mechanism responsible for the observed TR in the larger 100 nm GPG models that we observed.

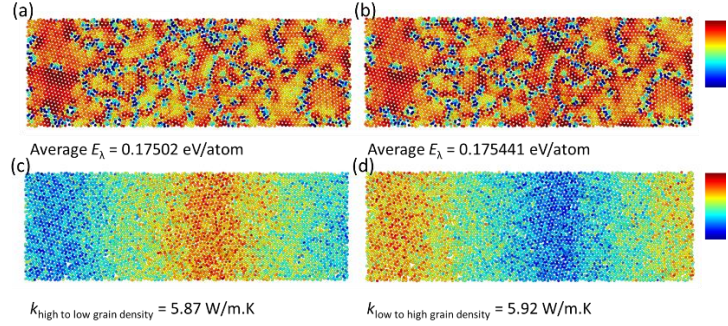

**FIG. 4** Map of spatial energy distribution, $E_\lambda$, in a 20 nm long GPG for opposite heat flow directions with respect to the grain density gradient is shown in (a) and (b) along with the corresponding per atom average temperatures in (c) and (d), respectively (the bounds of the color bars are 0.163 eV/atom at lower blue end, and 0.183 eV/atom for upper red end for (a) and (b), and 284 K at lower blue end, and 316 K for upper red end for (c) and (d)). The corresponding average $E_\lambda$ and $k$ values are mentioned below the map of the respective structure.

Thus, we investigate the second possible mechanism of TR for GPG, which depends on the varying temperature ($T$) sensitivities of $k$ for different regions in the 2D thermal rectifier[3]. We calculated the $k$ in single crystalline graphene and uniform grain sized polycrystalline graphene samples with (a) varying grain sizes and (b) varying temperatures and determined the variation in $k$, verifying that they follow a $\boldsymbol{k \propto T^{-\alpha}}$ relationship, where the value of the constant $\boldsymbol{\alpha}$ for a given uniform graphene structure indicates how strongly its $k$ varies with $T$ (Fig. 5. (a)). Fig. 5(b) shows that the $T$-sensitivity of $k$ is higher for larger grain sizes and reduces with reducing grain sizes, with the single crystalline graphene having the highest sensitivity of $k$ to $T$. Furthermore, the relationship between $k$ and $T$ for each uniform grained graphene structure clearly shows that the thermal conductivity of graphene decreases with increasing temperature. These observations imply that for GPG, the *thermal conductivity will be higher* when the larger grain size region (low $T$-sensitivity of $k$) of the GPG is at the colder end, i.e., the heat sink, and the smaller grain size region (low $T$-sensitivity of $k$) is at the hotter end, i.e., the heat source.

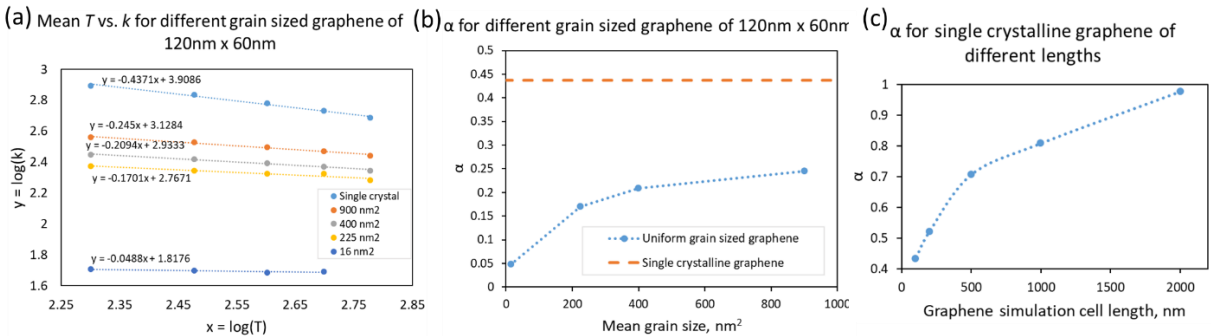

**FIG. 5.** Varying temperature dependence of thermal conductivities for graphene with different uniform grain sizes and for single crystalline graphene. (a) plots the log of $k$ versus log of the average $T$, which varies between 200 K to 600 K with 100 K intervals, for NEMD simulations of single crystal graphene and different uniform grain sized polycrystalline structures with the same total size but varying average grain

size. The linear fit equation for each structure is given close to the corresponding plot. (b) plots the changing values of $\alpha$ for the 4 different polycrystalline structures obtained from the linear fitting in (a) connected with a line for visual clarity, along with a dotted line denoting the value of $\alpha$ for single crystal graphene. (c) plots the variation of $\alpha$ for single crystalline graphene models versus different model lengths connected with a line for visual clarity.

It is understood that for pristine graphene, $k \propto T^{-1}$ due to the increased Umklapp scattering of long-range propagating phonons, which dominate the thermal transport in graphene, at higher temperatures[9, 23, 24]. Therefore, the increased localization of phonons in polycrystalline graphene with decreasing grain size (Fig. 3(a)) leads to a reduction of its $T$-sensitivity of $k$ as we observed. While the graphene models used for this comparative analysis between different grain sizes were of 120 nm in length and 60 nm in width each, we have also verified that the value of $\alpha$ for single crystalline graphene models of 7 nm unit width further increases with increasing length of the sample, converging to $\alpha = 1$ (Fig. 5(c)). Therefore, although the smaller model lengths result in smaller calculated values of $\alpha$ due to a reduced capacity to sample long range phonons in the atomistic simulations, the *qualitative nature of variation* of $\alpha$ for different grain sizes that we found should hold true irrespective of the sample size. In fact, the difference between the actual $T$-sensitivity of $k$ for different grain sizes in graphene should only become even more prominent in absolute terms for larger length samples. Hence, our results indicate that the grain-size dependent local $k$ of graphene could be the dominant mechanism which causes the observed direction of TR in GPG (Fig. 2(e)), such that the $k$ is higher when the heat flows from high grain density to low grain density, while the first mechanism based on the different propagating phonon energies at the heat source also acts on the GPG structures simultaneously with an opposite effect. This implies that the competitive interplay between the two TR mechanisms limits the TR value in GPG.

This hypothesis of the two competitive mechanisms of TR involved in GPG leads to some interesting conclusions which can be used to test the same. The way the higher energy propagating phonons generated at the heat source could increase the $k$ is by coupling across the length of the thermal rectifier device, resulting in an increase in the energies of propagating phonons even at the heat sink[3]. However, with increased length of the TR device, the long range phonons undergo more scattering while traveling from one end of the device to the other, which reduces the effect of the higher energy of those propagating phonons at the heat source[3]. Hence, the effect on $k$ of the first mechanism based on the propagating phonon energy at the heat source should get reduced with increasing device length, even when the difference in terminal grain densities remains the same[3].

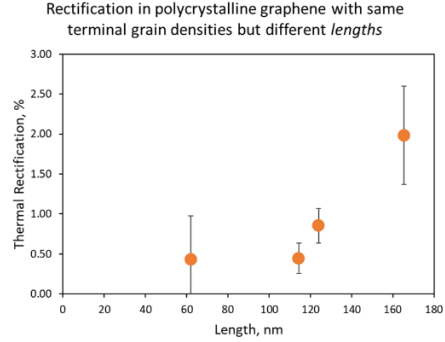

**FIG. 6.** Variation of average TR in GPG with varying lengths but same terminal grain densities (error bars denote the SD from multiple simulations).

To test this, we created four different GPG structures with varying lengths but the same terminal grain densities and a linear increase in the number of grains per width across the length. We calculated the TR for each using the same NEMD method. The results are plotted in Fig. 6 with the x-axis showing the length of the graded structures having the same width of 40 nm. Since the terminal grain densities are the same for all the structures, the gradient reduces with increasing length of the structure. It is clearly observed that the TR increases notably for increasing length, or decreasing grain density gradient, when the same terminal grain densities are maintained. This peculiar result aligns well with the explanation that with increasing length, the effect of the second mechanism of TR based on asymmetric $T$-sensitivity of $k$ remains the same while the effect of the competing first mechanism based on propagating phonon energy reduces, resulting in a higher dominance of the second mechanisms and consequently a net increase in the TR of the GPG models along the same direction of heat flow as the second mechanism.

**Conclusion**

This study provides the first evidence of the competing dual mechanisms that dictate the thermal rectification behavior in polycrystalline graphene, viz., asymmetric (i) propagating phonon coupling and (ii) temperature-dependence of thermal conductivity based on large-scale atomistic simulations and phonon calculations. This understanding can be vital in future efforts to manipulate the TR in practical polycrystalline graphene structures for more efficient 2D thermal rectifiers. Specifically, we expect engineered graded polycrystalline structures with a combination of greater length and larger difference between terminal grain densities to be promising candidates for efficient thermal rectifiers. Indeed, further structural modifications such as gradations in functionalization and defects can further enhance the rectification observed in similar graphene samples synergistically.

**Author information**

***Corresponding Author:** Simanta Lahkar – simanta.l@iitgn.ac.in

**Author Contributions:** SL conceptualized the work, executed and analyzed the simulations and wrote the manuscript. RR contributed towards conceptualization, analysis and revision of the manuscript. All authors have given approval to the final version of the manuscript.

**Competing interests statement**

The authors declare no competing financial or non-financial interests.

**Acknowledgement**

The authors acknowledge the Param Ananta computing facility at IIT Gandhinagar for all the simulations reported in this work. Funding from the Early Career Fellowship program at IIT Gandhinagar is also acknowledged.

**Data availability**

The typical simulation and analysis scripts, along with sample results and structures are available at https://github.com/simantalahkar/poly-tr-mechanism. Any other data related to this study will be provided by the corresponding author upon reasonable request.